\documentclass[conference]{IEEEtran}
\IEEEoverridecommandlockouts
\usepackage{float} 
\usepackage{cite}
\usepackage{amsmath,amssymb,amsfonts}
\usepackage{algorithmic}
\usepackage{graphicx}
\usepackage{textcomp}
\usepackage{xcolor}
\usepackage{subcaption}
\usepackage[section]{placeins} 
\def\BibTeX{{\rm B\kern-.05em{\sc i\kern-.025em b}\kern-.08em
    T\kern-.1667em\lower.7ex\hbox{E}\kern-.125emX}}
\begin{document}

\title{Path-Based Correlation Analysis of Meteorological Factors and eLoran Signal Delay Variations
\thanks{This work was supported in part by Grant RS-2024-00407003 from the ``Development of Advanced Technology for Terrestrial Radionavigation System'' project, funded by the Ministry of Oceans and Fisheries, Republic of Korea;
in part by the National Research Foundation of Korea (NRF), funded by the Korean government (Ministry of Science and ICT), under Grant RS-2024-00358298; 
in part by the Future Space Navigation and Satellite Research Center through the NRF, funded by the Ministry of Science and ICT (MSIT), Republic of Korea, under Grant 2022M1A3C2074404; 
and in part by the MSIT, Korea, under the Information Technology Research Center (ITRC) support program supervised by the Institute of Information \& Communications Technology Planning \& Evaluation (IITP) under Grant IITP-2024-RS-2024-00437494.
}
}

\author{\IEEEauthorblockN{Junwoo Song} 
\IEEEauthorblockA{\textit{School of Integrated Technology} \\
\textit{Yonsei University}\\
Incheon, Korea \\
junwoo3529@yonsei.ac.kr} 
\and
\IEEEauthorblockN{Pyo-Woong Son${}^{*}$} 
\IEEEauthorblockA{\textit{Department of Electronics Engineering} \\
\textit{Chungbuk National University} \\
Cheongju, Korea \\
pwson@cbnu.ac.kr}
{\small${}^{*}$ Corresponding author}
}

\maketitle

\begin{abstract}
Unlike GNSS, which is vulnerable to jamming and spoofing due to its inherently weak received power, eLoran exhibits robustness owing to its high field strength.
Therefore, the eLoran system can maintain reliable operation even in scenarios where GNSS becomes unavailable. 
However, since eLoran signals propagate through ground waves, the propagation delay is susceptible to changes in surface conditions, including both terrain and meteorological variations. 
This study aims to analyze the correlation between the temporal variations in eLoran signal propagation delay and meteorological factors at various points along the signal path.
\end{abstract}


\section{Introduction}
Global Navigation Satellite Systems (GNSS) \cite{Kim14:Comprehensive, Chen11:Real, Lee23:Seamless, Kim23:Machine, Lee24:A, Kim23:Single, Kim22:Machine, Lee22:Urban, Lee22:Optimal, Kim19:Mitigation, Jeong24:Quantum, Kim25:Set} are indispensable technologies that serve as the backbone of numerous applications in the military, transportation, and logistics domains. 
Despite widespread adoption, GNSS signals are inherently weak and thus highly susceptible to jamming and spoofing\cite{Park21:Single, Park17:Adaptive, Moon24:HELPS, Kim23:Low, Lee22:Evaluation, Park25:Toward, Rhee21:Enhanced}. 
In fact, South Korea has experienced service disruptions caused by deliberate jamming attacks from North Korea, highlighting the vulnerability of GNSS\cite{Kim22:First, Son20:eLoran}.

To address these limitations, the enhanced Loran (eLoran) system has emerged as an advanced terrestrial navigation technology\cite{Son22:Compensation}.
eLoran operates at a low frequency of 100 kHz, transmitting signals as ground waves, and has a significantly higher received field strength compared to GNSS\cite{Yin25:Preliminary}. 
Owing to its high field strength, eLoran is more resistant to signal disruption and interference than GNSS.
As such, eLoran serves as a promising complementary or backup Positioning, Navigation, and Timing (PNT) solution in the event of GNSS failure\cite{Yang25:Research}.
In South Korea, the eLoran infrastructure was implemented between 2016 and 2020 and currently consists of a central control station, three transmitting stations located in Pohang, Gwangju, and Socheong Island, and two reference stations in Incheon and Pyeongtaek\cite{Son23:Demonstration, Son22:Compensation}.

Unlike GNSS, which transmits signals from satellites, eLoran signals are transmitted from ground-based stations and propagate along the Earth's surface, including terrain and sea surfaces\cite{Cheng25:Research}.
In ground-wave propagation, signal delay is primarily caused by the electromagnetic properties of the transmission medium, including its conductivity and permittivity\cite{Tao2025:Ameteorological, Xuyin24:Prediction, Hehenkamp2023:Prediction}.
Accordingly, accurate estimation and compensation of this signal propagation delay are crucial for improving the PNT performance of the eLoran system.
In light of this, various efforts have been made to estimate and mitigate propagation delay in eLoran systems\cite{Son24:eLoran,  Son23:Demonstration, Son22:Compensation}.
One such approach is the use of ASF (Additional Secondary Factor) maps, which are designed to capture spatial variations in propagation delay and play a central role in delay correction in eLoran systems.
However, these maps often fail to account for temporal variations in the delay observed at the same location over time.
To address this limitation, several studies have focused on characterizing the time-varying nature of propagation delay under dynamic atmospheric and ground conditions\cite{Kang25:Enhancing, Tao2025:Ameteorological, Xuyin24:Prediction, Di25:Comparative, Di2025:longwave}.
These studies have demonstrated that temporal variations in meteorological conditions can influence the state of the propagation medium, leading to changes in signal propagation delay.
As eLoran signals travel along the Earth's surface, the propagation delay is influenced more by the environmental conditions along the signal path than by those at the transmitter or receiver alone.

In this study, we investigate the correlation between temporally varying eLoran signal propagation delays and the meteorological factors along the transmission path.
The signal path is spatially sampled, and meteorological data for each sampling point are interpolated from nearby weather observation stations using the inverse distance weighting (IDW) method.
The propagation delay is derived from Time of Arrival (TOA) measurements collected at an eLoran receiver.
The objective is to reveal how environmental factors distributed along the propagation path influence the delay characteristics of the eLoran system. To quantify the correlation, Pearson correlation coefficients are computed between the interpolated meteorological factors and the observed propagation delays.

\section{Methodology}
\subsection{TOA-Based Delay Measurement}

The Time of Arrival (TOA) value measured at the receiver consists of two main components: the signal propagation time and the internal processing delay of the receiver\cite{Tao2025:Ameteorological}. 
The processing delay is assumed approximately constant over the experiment, being determined by the receiver hardware and signal-processing pipeline.
Accordingly, we attribute temporal variations in \text{TOA} primarily to changes in signal propagation time.
Ideally, when both the transmitting and receiving stations are fixed, the measured TOA value remains constant. 
However, in practice, even with fixed stations, the measured TOA value varies over time.

\begin{equation}
 \text{TOA} = t_p + t_r
 \label{eqn:eLoran signal TOA equation}
\end{equation}

where $t_p$ denotes the signal propagation time from the transmitter to the receiver, and $t_r$ represents the internal processing delay of the receiver.

The signal propagation time $t_p$ can be decomposed as follows:

\begin{equation}
 t_p= PF + SF + ASF
 \label{eqn:propagation_delay_components}
\end{equation}

where $PF$ denotes the Primary Factor, $SF$ the Secondary Factor, and $ASF$ the Additional Secondary Factor.
According to \cite{Tao2025:Ameteorological, Xuyin24:Prediction}, $PF$ is calculated as follows:

\begin{equation}
 \mathrm{PF} = \ \int \frac{n_s}{c} \, ds
 \label{eqn:PF calculation}
\end{equation}

where $n_s$ denotes the atmospheric refractive index along the signal path, $c$ is the speed of light in a vacuum, and $s$ represents the signal propagation path.
The atmospheric refractive index $n$ is determined by temperature, atmospheric pressure and water vapor pressure, and can be calculated as follows.

\begin{equation}
N = (n - 1) \times 10^6 \approx \frac{77.6 \, P}{T} + \frac{3.73 \times 10^5 \, e_w}{T^2}
\label{eqn:refractive_index}
\end{equation}

where $T$ denotes the air temperature, $P$ is the atmospheric pressure, and $e_w$ represents the water vapor pressure.
$SF$ refers to the delay caused by signal propagation over seawater, while $ASF$ represents the delay incurred over land. 
Although the propagation environments differ, both delay components are determined based on the same physical principle, which involves the phase attenuation factor $W$.
According to \cite{Tao2025:Ameteorological, Xuyin24:Prediction}, $SF$ is calculated as follows:

\begin{equation}
    SF = \frac{10^6}{\omega} \arg(W)
    \label{eqn:SF_equation}
\end{equation}

where $\omega$ denotes the central angular frequency, and $W$ represents the phase attenuation factor.
Here, $W$ is determined by the electrical conductivity and permittivity of the medium. 

Therefore, changes in the electromagnetic properties of the medium, caused by meteorological variations along the propagation path, are expected to affect $W$, which governs both $SF$ and $ASF$.
Also, meteorological conditions influence the atmospheric refractive index $n$, which in turn affects $PF$. Consequently, changes in meteorological factors lead to variations in all these delay components, ultimately resulting in fluctuations of the measured \text{TOA}.

\subsection{Data Collection and Processing}

\begin{figure}[H]
    \centering
    \includegraphics[width=0.8\linewidth]{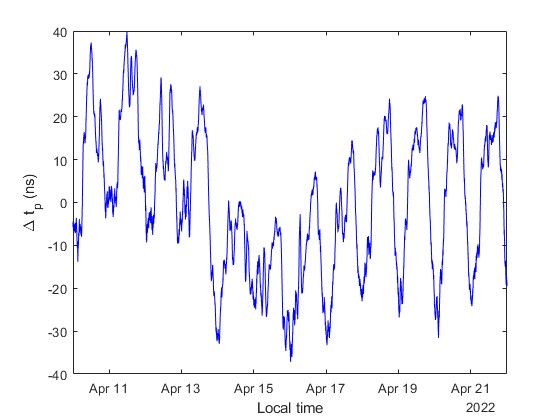}
    \caption{Processed TOA measurements over a 12-day period in April.}
    \label{fig:TOA}
\end{figure}

The data used in this study consist of two types: \text{TOA} measurements obtained from an eLoran receiver and meteorological observations collected from weather stations. 
The \text{TOA} measurements were obtained over the course of April from eLoran signals transmitted by the eLoran transmitter (9930W) located in Gwangju and received by a receiver located in Pyeongtaek, at a distance of approximately 218 km.
The collected TOA data were processed by first removing outliers attributable to measurement errors, then subtracting the mean, and finally applying a 1-hour moving-average filter. The temporal variation, $\Delta\text{TOA}$, is defined as in (\ref{eq:delta_tp}).
Fig.~\ref{fig:TOA} illustrates the temporal variation of $\Delta \text{TOA}$.

\begin{equation}
\Delta \text{TOA} = \text{TOA} - \overline{\text{TOA}}
\label{eq:delta_tp}
\end{equation}
where $\overline{\mathrm{TOA}}$ denotes the mean value of Time of arrival.

To generate multiple analysis points along the signal propagation path, we sampled the path at equal spatial intervals.
Meteorological observations to be used for interpolation were collected from ten weather stations, all located within 30\,km of the path.
Subsequently, the observations were interpolated to the sampling points along the path using inverse distance weighting (IDW), yielding interpolated estimates of the meteorological factors.

\begin{equation}
\hat{z}(x_0) = \frac{\sum_{i=1}^N w_i z(x_i)}{\sum_{i=1}^N w_i}, 
\quad \text{where} \quad
w_i = \frac{1}{d(x_0, x_i)^p}
    \label{eqn:IDW}
\end{equation}
where $\hat{z}(x_0)$ is the estimated value at location $x_0$, $z(x_i)$ is the observed value at weather station $x_i$, $d(x_0, x_i)$ denotes the distance between $x_0$ and $x_i$, and $p$ is the weighting power parameter, which was set to 2 in this study.

\subsection{Correlation Analysis between Meteorological Factors and \text{TOA} Variations}

This study aims to analyze the correlation between meteorological data at multiple points along the signal propagation path and $\Delta \text{TOA}$.
To this end, the Pearson correlation coefficient $r$ was used as a metric for the correlation analysis. 
This coefficient quantifies the degree of linear relationship between two variables and is defined as

\begin{equation}
    r = \frac{\sum_{i=1}^{n} (a_i - \bar{a})(b_i - \bar{b})}
             {\sqrt{\sum_{i=1}^{n} (a_i - \bar{a})^2} \, \sqrt{\sum_{i=1}^{n} (b_i - \bar{b})^2}}
    \label{eqn:pearson correlation coefficient}
\end{equation}
where $a_i$ and $b_i$ are the paired samples, and $\bar{a}$ and $\bar{b}$ are their respective mean values.  
A correlation coefficient close to $+1$ indicates a strong positive correlation, while a value close to $-1$ indicates a strong negative correlation. 
In contrast, if the correlation coefficient is close to $0$, it implies that there is no linear relationship between the two variables.
\section{Results}
\subsection{Correlation Analysis at a Single point}
\begin{figure}
    \centering
    % Row 1
    \begin{subfigure}{0.48\linewidth}
        \centering
        \includegraphics[width=\linewidth]{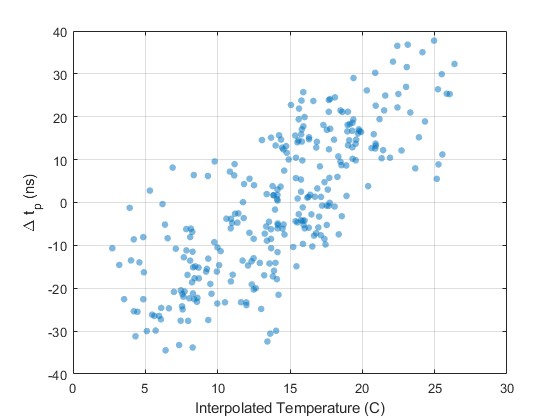}
        \caption{($r=0.7718$)}
        \label{fig:temp}
    \end{subfigure}
    \hfill
    \begin{subfigure}{0.48\linewidth}
        \centering
        \includegraphics[width=\linewidth]{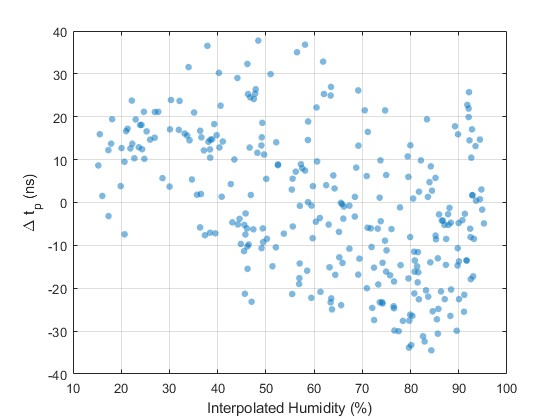}
        \caption{($r=-0.4305$)}
        \label{fig:hum}
    \end{subfigure}

    \vspace{0.4cm}

    % Row 2
    \begin{subfigure}{0.48\linewidth}
        \centering
        \includegraphics[width=\linewidth]{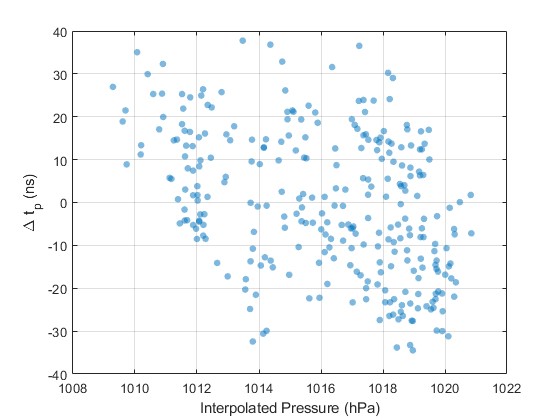}
        \caption{($r=-0.4524$)}
        \label{fig:press}
    \end{subfigure}
    \hfill
    \begin{subfigure}{0.48\linewidth}
        \centering
        \includegraphics[width=\linewidth]{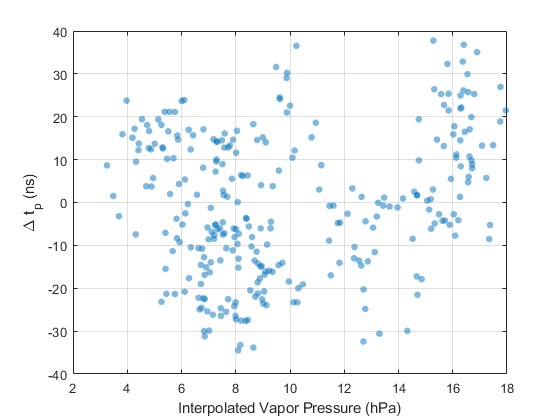}
        \caption{($r=0.2925$)}
        \label{fig:vpress}
    \end{subfigure}

    \vspace{0.4cm}

    % Row 3
    \begin{subfigure}{0.48\linewidth}
        \centering
        \includegraphics[width=\linewidth]{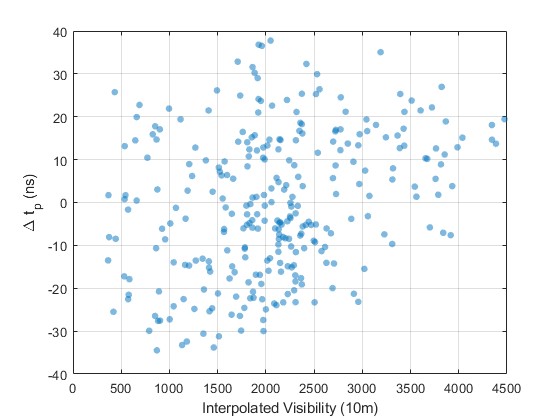}
        \caption{($r=0.4165$)}
        \label{fig:vis}
    \end{subfigure}
    \hfill
    \begin{subfigure}{0.48\linewidth}
        \centering
        \includegraphics[width=\linewidth]{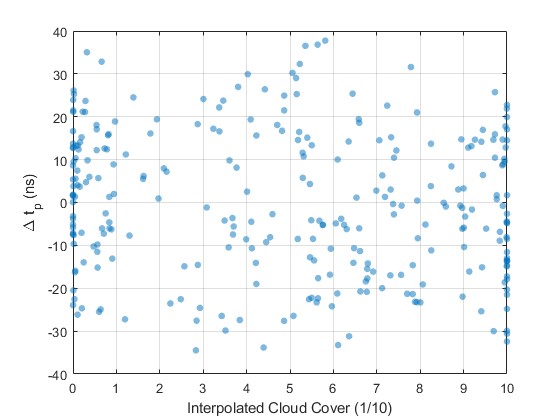}
        \caption{($r=-0.1403$)}
        \label{fig:cloud}
    \end{subfigure}

    \caption{Scatter plots between each meteorological factor at the receiver location and $\Delta \text{TOA}$. (a)–(f) represent temperature, humidity, pressure, vapor pressure, visibility, and cloud amount, respectively.}
    \label{fig:multi_vars}
\end{figure}

To analyze the relationship between meteorological factors and $\Delta \text{TOA}$ at a single observation point, scatter plots were used for visualization.
In Fig. \ref{fig:multi_vars}, the x-axis represents the meteorological factors at the sampling point corresponding to the receiving location, and the y-axis indicates $\Delta \text{TOA}$.
In Fig.~\ref{fig:multi_vars}(\subref{fig:temp}), temperature exhibits a strong positive linear correlation with $\Delta \text{TOA}$. 
By contrast, humidity and pressure in Fig.~\ref{fig:multi_vars}(\subref{fig:hum}) and Fig.~\ref{fig:multi_vars}(\subref{fig:press}) each show a weak negative correlation. 
Visibility in Fig.~\ref{fig:multi_vars}(\subref{fig:vis}) displays a weak positive association, whereas vapor pressure and cloud amount in Fig.~\ref{fig:multi_vars}(\subref{fig:vpress}) and Fig.~\ref{fig:multi_vars}(\subref{fig:cloud}) exhibit no clear linear relationship.
Notably, the captioned Pearson $r$ values for Fig. \ref{fig:multi_vars}(a)-(f) numerically corroborate the qualitative patterns described above.

\subsection{Correlation Analysis Across Multiple Sampling Points}

Beyond analyzing the correlation between meteorological factors and $\Delta \text{TOA}$ at a single point, we evaluated correlations at multiple points along the propagation path to assess spatial heterogeneity in these relationships. 
Specifically, we extended the analysis to 50 uniformly sampled points along the path and, at each point, computed the Pearson correlation coefficient between each meteorological factor and $\Delta \text{TOA}$.

\begin{figure}
\centering
    \includegraphics[width=1.0\linewidth]{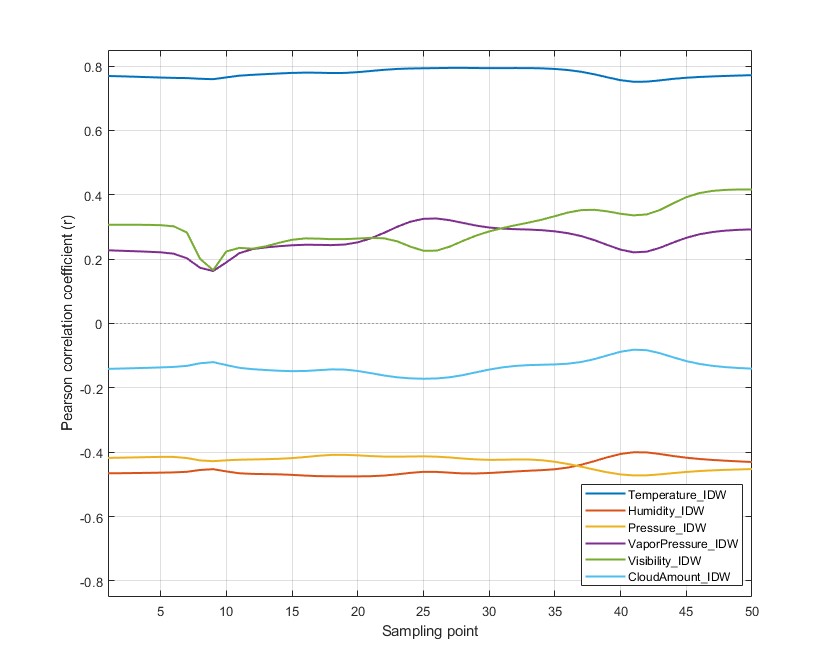}
\caption{Variation of Pearson correlation coefficients between each meteorological factor and $\Delta \text{TOA}$ along the signal propagation path.  Sampling point 1 corresponds to the transmitting station, and sampling point 50 corresponds to the receiving station.}
\label{fig:correlation_graph}
\end{figure}

Fig. \ref{fig:correlation_graph} shows the correlation coefficients for each meteorological factors at all sampling points, and Table \ref{tab:correlation_table} presents their mean and standard deviation.
In general, the across-path mean is broadly consistent with the coefficient at the receiving point for most factors; an exception is visibility, whose coefficient at the receiving point differs substantially from the across-path mean.
The standard deviations together with the spatial profiles in Fig. \ref{fig:correlation_graph} indicate that temperature, humidity, and pressure exhibit relatively strong linear correlations that vary little across  points.
In contrast, visibility and vapor pressure show larger standard deviations and pronounced spatial fluctuations, suggesting localized peaks and troughs rather than a stable correlation with $\Delta \text{TOA}$.
Cloud amount shows no linear association with $\Delta \text{TOA}$ at any sampling point, resulting in a small standard deviation of the coefficients.

Local meteorological conditions differ at every sampling point. Accordingly, the calculated correlation coefficients vary spatially. In particular, for meteorological factors whose temporal variations differ markedly across segments of the propagation path, the resulting correlations with $\Delta\mathrm{TOA}$ vary appreciably by location rather than remaining uniform.

\begin{table}
\centering
\caption{Mean and standard deviation} of Pearson correlation coefficients for each meteorological factor, calculated over all points along the propagation path.
\begin{tabular}{lcccccc}
\hline
 & Tem & Hum & Pres & Vap Pres & Vis & Cloud Amt \\
\hline
Mean & 0.776 & -0.452 & -0.430 & 0.258 & 0.299 & -0.135 \\
Std & 0.013 & 0.023 & 0.020 & 0.039 & 0.601 & 0.022 \\

\hline
\end{tabular}
\label{tab:correlation_table}
\end{table}

\section{Conclusion}

This study investigated the influence of meteorological factors along the signal propagation path on variations in Time of Arrival (\text{TOA}) measurements. 
Meteorological data along the path were generated using the inverse distance weighting (\text{IDW}) interpolation method, and Pearson correlation coefficients were computed to evaluate the correlations between each meteorological factors and \text{TOA} variations at each sampling point.

The results revealed both the relationships between individual meteorological factors and \text{TOA} variations, as well as their spatial variability along the propagation path.
Specifically, temperature, humidity, and pressure exhibited relatively strong and spatially stable correlations with \text{TOA} variations, whereas visibility and vapor pressure showed pronounced segment-dependent fluctuations, and cloud amount displayed little to no linear correlation.

This study has limitations in that the meteorological data were not directly measured on the actual path but interpolated from nearby stations. 
Furthermore, Pearson correlation captures only linear dependencies, limiting the ability to identify nonlinear or complex interactions.

Future work will focus on quantitatively assessing the differences between interpolated and directly observed meteorological data, and applying methods capable of capturing nonlinear and complex relationships between meteorological factors and \text{TOA} variations.

\section*{Acknowledgment}

Generative AI (ChatGPT, OpenAI) was used solely to assist with grammar and language improvements during the manuscript preparation process.  
No content, ideas, data, or citations were generated by AI.  
All technical content, methodology, analysis, and conclusions were written and verified solely by the authors.

\bibliographystyle{IEEEtran}
\bibliography{mybibfile, IUS_publications}

\vspace{12pt}

\end{document}